\documentclass[aps,prb,twocolumn,groupedaddress,amssymb,floatfix]{revtex4}

\usepackage{graphicx}
\usepackage{dcolumn}

\begin{document}

\preprint{}

\title{Magnetization of a two-dimensional electron gas with a second filled subband}

\author{M.R. Schaapman}
\email[]{maaikes@sci.kun.nl}
\author{U. Zeitler}
\author{P.C.M. Christianen}
\author{J.C. Maan}
\affiliation{Nijmegen Science Research Institute of Matter, High
Field Magnet Laboratory, University of Nijmegen, Toernooiveld 7,
6525 ED Nijmegen, The Netherlands}

\author{D. Reuter}
\author{A.D. Wieck}
\affiliation{Lehrstuhl f\"ur Angewandte Festk\"orperphysik,
Ruhr-Universit\"at Bochum, Universit\"atsstrasse 150, 44780
Bochum, Germany}

\author{D. Schuh}
\author{M. Bichler}
\affiliation{Walter Schottky Institut, Technische Universit\"at
M\"unchen, Am Coulomb Wall, 85748 Garching, Germany}

\date{\today}

\begin{abstract}
We have measured the magnetization of a dual-subband
two-dimensional electron gas, confined in a GaAs/AlGaAs
heterojunction. In contrast to two-dimensional electron gases with
a single subband, we observe non-$1/B$-periodic, triangularly
shaped oscillations of the magnetization with an amplitude
significantly less than $1~\mu_{\mathrm{B}}^*$ per electron. All
three effects are explained by a field dependent self-consistent
model, demonstrating the shape of the magnetization is dominated
by oscillations in the confining potential. Additionally, at 1~K,
we observe small oscillations at magnetic fields where
Landau-levels of the two different subbands cross.
\end{abstract}


\maketitle


When an extra degree of freedom is added to a two-dimensional
electron gas (2DEG), many-body interactions can lead to the
formation of novel electronic grounds-states at the crossings of
the different energy-levels in the system.~\cite{DasSarma1}
Two-dimensional electron gases with crossing energy levels can be
realized in a variety of systems with different relative sizes of
orbital and spin effects, Coulomb energy, and different coupling
between the components.  Their study has lead to the discovery of
many correlated Quantum Hall
states,~\cite{Piazza1,Poortere1,Jaroszynski1} and much effort is
put into unravelling the energy-level structure of these systems.

One way of realizing such a 2D system, is to increase the electron
density of a III-V 2DEG such that a second subband becomes
occupied. Dual-subband systems realized in a quantum well have
recently been studied within this context.~\cite{Pan1,Muraki1} A
similar system is the dual-subband 2DEG in a GaAs/AlGaAs
heterojunction. In transport studies, the multi-subband 2DEG is
generally assumed to be a superposition of single 2DEGs:  the
Landau-level structure is a superposition of Landau fans,
separated by the inter-subband spacing calculated
self-consistently at zero magnetic field.~\cite{Ando4,Stern1}

In this paper we study the magnetization ($M$) of a dual-subband
2DEG.  This is a way to directly probe a thermodynamic property,
the chemical potential, $\mu$. For two-dimensional systems, the
Maxwell relation between $M$ and $\mu$ is reduced to a
proportionality. Since the Fermi energy ($E_\mathrm{F}$) is equal
to $\mu$ at low temperatures, the magnetization directly reveals
changes in the size as well as the shape of the Fermi energy:
$\Delta M=\frac{N}{B}\Delta E_\mathrm{F} \label{eq:Max}$, where
$N$ is the total number of electrons.

The magnetization of multi-subband 2DEGs has already attracted
some attention, both theoretically~\cite{Alexandrov1,Alexandrov2}
and experimentally,~\cite{Shepherd} however, these studies have
focussed on very high density systems with three or more filled
subbands. In this regime changes in the energy-gap between the
subbands can be ignored.  In this paper we focus on the effect of
the filling of a only a second electronic subband on the Fermi
energy.

Quantum oscillations in the magnetization of a single 2DEG are
well known to be characterized by strictly $1/B$-periodic
saw-tooth-like oscillations with an amplitude of one effective
Bohr magneton ($\mu_{\mathrm{B}}^*$) per
electron.~\cite{Schoenberg, Eisenstein1, Wiegers5}  Here we will
show that this is no longer the case in a multi-component system.
Due to a self-consistent, magnetic field dependent redistribution
of electrons between the subbands inside the heterojunction, the
amplitude of the oscillations becomes considerably reduced, the
saw-tooth-like steps are broadened into triangles, and the
$1/B$-periodicity is lost. Additionally we find that extra
magnetization minima appear at low temperature at the Landau-level
crossings of the two subbands.


We study the magnetization of two samples with different electron
densities, realized in GaAs/AlGaAs heterojunctions grown by
molecular beam epitaxy: a single subband 2DEG (sample 1) and a
high density 2DEG (sample 2) where two electronic subbands are
occupied. Sample 1 has a density of $4.8 \cdot 10^{11}$~cm$^{-2}$,
and a mobility of $2.2 \cdot 10^{6}$~cm$^2$/Vs. Our high electron
density sample 2 has a carrier concentration of $8.0 \cdot
10^{11}$~cm$^{-2}$ and a mobility of $1.4 \cdot 10^{6}$~cm$^2$/Vs.
Most of the electrons ($7.4 \cdot 10^{11}$~cm$^{2}$, deduced from
transport measurements on a reference sample) remain in the lowest
subband; the small remaining fraction occupies the second subband.
The magnetization experiments were performed using a torsional
magnetometer with optical angular detection.~\cite{Schaapman}

\begin{figure}[t]
\includegraphics[scale=0.5]{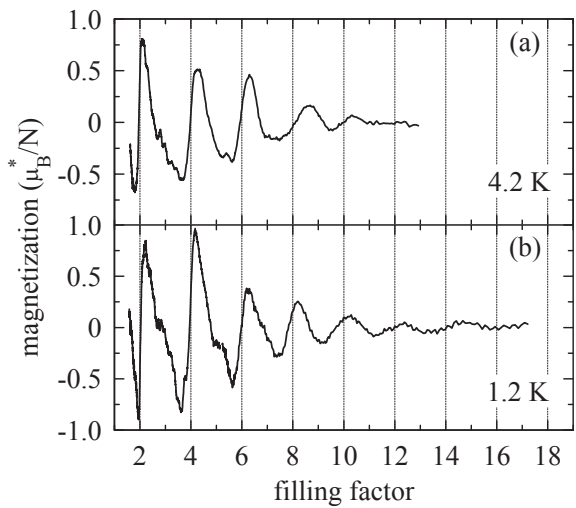}
\caption{Magnetization of sample 1 (single 2DEG with $n=4.8 \cdot
10^{11}$ cm$^{-2}$) at 4.2~K (a) and 1.2~K (b).  Oscillations are
a strictly $1/B$-periodic saw-tooth, the amplitude saturates to
$1~\mu_{\mathrm{B}}^*$ per electron at low $\nu$.  Features at
filling factors 3 and 5 are related to
spin-splitting.\label{fig:single}}
\end{figure}

Figure~\ref{fig:single} shows the magnetization of sample 1, the
single subband 2DEG, as a function of filling factor
($\nu=hn/eB$).  It displays oscillations periodic in $\nu$, i.e.
$1/B$-periodic oscillations.  The steps at the lowest filling
factors, where Landau-level broadening has the least influence,
are saw-tooth shaped, and, with decreasing temperature, the
amplitude saturates to $1~\mu_{\mathrm{B}}^*$ per electron.
Although the steps are rather sharp, even at 1.2~K they still have
a small, finite width indicating a finite density of states (DOS)
in between Landau-levels.~\cite{Wiegers5,Schwarz4}

Apart from the clear steps assigned to the Landau gap at even
integer filling factors, at 1.2~K (figure~\ref{fig:single}(b))
additional features appear at odd integer filling factors. They
are attributed to the opening of a spin-gap, significantly
enhanced due to exchange
interaction.~\cite{Nicholas1,Grundler1,Meinel2}

\begin{figure}[t]
\includegraphics[scale=0.5]{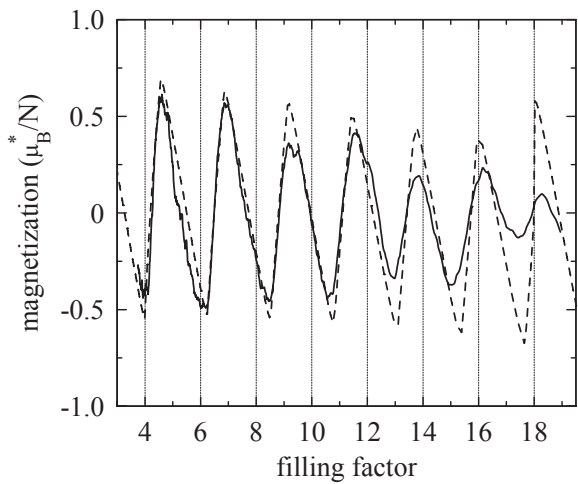} 
\caption{Magnetization of the dual-subband 2DEG with $n=8.0 \cdot
10^{11}$ cm$^{-2}$ at 4.2~K. The dashed line is a self-consistent
calculation using a Gaussian Landau-level broadening with
$\Gamma=0.2\sqrt{B}$~meV.  Note the deviation from
$1/B$-periodicity.\label{fig:multi}}
\end{figure}

\begin{figure}[t]
\includegraphics[scale=0.5]{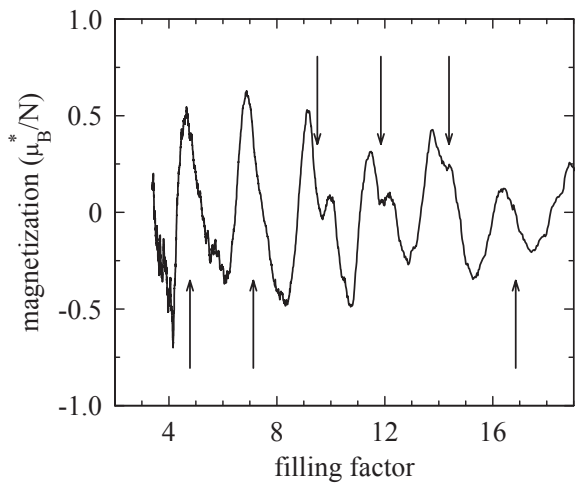}
\caption{Magnetization of a dual-subband 2DEG at 1~K.  Arrows
indicate the positions of Landau-level crossings, additional
features can be seen at these positions in intermediate magnetic
fields. \label{fig:minima}}
\end{figure}

The second sample, with two filled subbands, displays the
magnetization plotted by the solid line in figure~\ref{fig:multi}.
The dashed line is a theoretical calculation and will be discussed
later. Also for this sample the magnetization oscillates as a
function of inverse magnetic field. Closer inspection of the data,
however, reveals three distinct differences compared to the single
2DEG. First, the oscillations are no longer sawtooth-like, but
instead they are triangularly shaped. Secondly, we find that the
oscillations are no longer strictly periodic in $1/B$.  This
becomes clear when we see that while $\nu=4$ coincides with an
oscillation minimum, there is an increasing discrepancy, and
$\nu=14$ actually coincides with an oscillation maximum. Finally,
the amplitude of the oscillation is about $0.5~\mu_{\mathrm{B}}^*$
per electron, even for the lowest filling factors.  This value is
significantly less than the $1~\mu_{\mathrm{B}}^*$ per electron
observed in figure~\ref{fig:single}b for the single 2DEG and it
remains at this level even for the lowest temperatures (see
fig.~\ref{fig:minima}).

In order to understand the behavior of the magnetization, it is
important to realize that in a heterojunction the confining
potential of the 2DEG is formed by the electrons themselves. In a
dual-subband 2DEG redistribution of charge over the two subbands
can occur when a magnetic field is applied, resulting in a
potential that is not fixed as a function of magnetic field. The
wavefunction of the second subband is much more extended than that
of the first one, therefore even a small change in its occupation
can have profound effects. Since the occupation of the two
subbands depends on the magnetic field that quantizes the DOS into
Landau levels, and since the shape of the confining potential, the
inter-subband spacing and the spacial charge distribution are
interdependent, the Schr\"odinger and Poisson equations have to be
solved self-consistently for each value of the magnetic
field.~\cite{Sanchez,Trott}

In our modelling we keep the electron density fixed and assume a
Gaussian broadened DOS with a width that increases with the square
root of the magnetic field.  As the (bare) spin-splitting is too
small to have an effect, it is neglected in the calculations. The
Landau-level broadening is our, albeit only, fit-parameter. We
note that the calculated amplitude decreases with increasing
broadening, however, even in the limit of non broadened Landau
levels, the calculated amplitude of the oscillations is only
$0.7~\mu_{\mathrm{B}}^*$ per electron.

Using the self-consistent model, we have calculated the Fermi
energy as a function of magnetic field, from which the
magnetization follows directly through the Maxwell
proportionality. The dashed line in figure~\ref{fig:multi} shows
the resulting magnetization for a Gaussian Landau-level broadening
with $\Gamma=0.2\sqrt{B}$~meV. It is in very good agreement with
the experimental data as it reproduces all three observed effects:
triangular shape, non-periodicity, and the reduced amplitude.

\begin{figure}[t]
\includegraphics[scale=0.5]{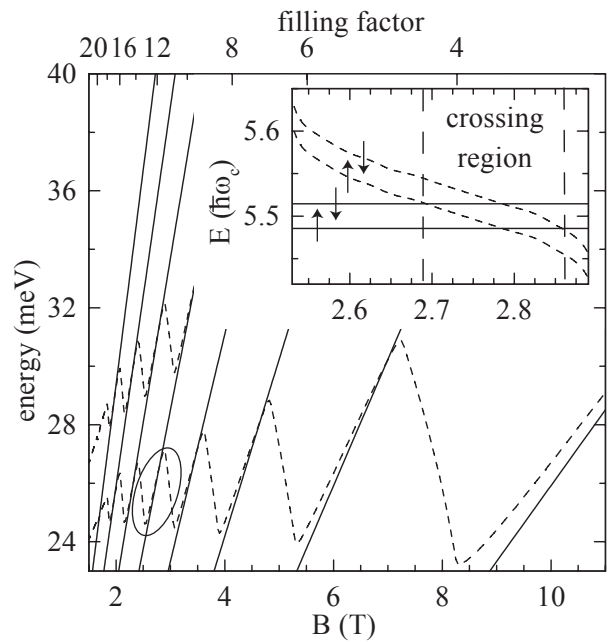}
\caption{Landau-level diagram of the dual-subband 2DEG. Solid
lines depict levels originating from the lower subband.  Dashed
lines show the Landau-levels of the higher subband. The inset is
an enlargement of the circled Landau-level crossing around
$\nu=11.9$, normalized to the cyclotron energy.\label{fig:LL}}
\end{figure}

Inspection of the self-consistent field-dependent modelling of the
high-density 2DEG in detail reveals two important points.  First,
although the number of electrons in the highest subband is small,
it remains populated up to high magnetic fields. Secondly, the
shape of the Fermi energy (and thus magnetization) is determined
by the oscillations in the inter-subband spacing, caused by
self-consistent redistribution of electrons over the two subbands.
A Landau-level scheme for  the dual-subband 2DEG, resulting from
the self-consistent model, is depicted in figure~\ref{fig:LL}.
While levels originating in the lower subband (solid lines) are
linear functions of the magnetic field, the Landau-levels of the
higher subband (dashed lines) oscillate according to the
inter-subband spacing. Above 1.5~T only the lowest Landau-level of
the second subband is populated and the Fermi energy lies
continuously within it.  It can be clearly seen that the
oscillations of the confining potential due to the redistribution
of the electrons have a large effect on the energy-level
structure.

At this stage it is interesting to remark that consequently the
wideness of the magnetization step is mainly caused by this
redistribution and not by a finite DOS between two Landau-levels
as suggested for a single subband 2DEG.~\cite{Wiegers5,Schwarz4}
Only in the region nearing $\nu=4$ there is, scarcely visible in
figure~\ref{fig:multi}, a kink followed by a sharp step, whose
finite width is related to this small, extra DOS.  These features
are not at all visible on the other downward slopes, where the
width is determined by the electron redistribution, and including
the extra DOS does not influence the shape of the calculated
Fermi-energy.

When reducing the temperature to 1~K, additional minima appear in
the 2DEG magnetization (figure~\ref{fig:minima}) around filling
factor $\nu=9.6$ and $\nu=12.0$, and $\nu=14.2$. Interestingly
enough these filling factors coincide with positions where two
Landau levels originating from the two subbands cross, indicated
by down-arrows in fig.~\ref{fig:minima}.

On the flanks of the triangular oscillations a series of crossings
occurs between the lowest Landau-level of the higher subband and
Landau-levels with decreasing index of the lower subband as the
magnetic field increases (see figure~\ref{fig:LL}). The addition
of spin-splitting to this (single-electron) picture results in
energy-level schemes as depicted in the inset of fig.~\ref{fig:LL}
for one of the crossings.  When spin-splitting is taken into
account, there is not a single level-crossing, but a small region
where the levels with different spin consecutively cross each
other.  In this region spin up and spin down do not alternate for
increasing energy:  the two spin down levels are lowest in energy,
the spin up levels the highest.

Although the Landau levels in figure~\ref{fig:LL} are represented
by discrete lines, they are in fact considerably broadened,
creating an overlap and giving the electrons some freedom to
distribute themselves over the available energy-levels. We suggest
this enables electrons to form a novel electronic ground-state
that is spin-polarized in the crossing region. Creation of this
polarized state would be favored by the system, because exchange
interaction significantly reduces the ground-state energy. When
the energy gain exceeds the broadening of the energy-levels, the
enhanced gap shows up as a minimum in the magnetization.

Although it is evident that at lower magnetic fields (filling
factors higher than 14, up-arrows indicate the positions of the
level crossings), extra structure cannot be seen due to the
broadness of the Landau-levels, extra structure is also too small
to be observed at the Landau-level crossings of the lowest filling
factors (up-arrows in fig.~\ref{fig:minima}).  Clearly the picture
of the crossing region sketched above is not yet complete and
further experimental and theoretical investigation of this
many-body effect is required.

In summary, we have measured the magnetization of the coupled
2DEGs in a dual-subband 2DEG in a GaAs/AlGaAs heterojunction. We
find that the de Haas-van Alphen oscillations are changed in three
ways compared to those of the single 2DEG. The shape is
triangular, the oscillation amplitude is reduced to
$0.5~\mu_{\mathrm{B}}^*$ and the oscillations are no longer
periodic in $1/B$. This behavior is well described by a
self-consistent model, taking into account changes of the
confining potential with magnetic field.  It shows the shape of
the Fermi energy and consequently the magnetization is entirely
dominated by the oscillations in this potential due to
redistribution of electrons over the two subbands. We observe
additional magnetization minima at 1~K, which occur at magnetic
fields corresponding to the positions where Landau-levels
originating in the two different subbands cross. These minima
possibly originate from a reduction of the total energy by the
formation of a novel, exchange enhanced electronic state at the
level-crossing.

\begin{acknowledgments}
We would like to thank M. Elliot for fruitful discussion.
\end{acknowledgments}

\bibliography{multiple,magnetometer,general,Qhfm}

\end{document}